\documentclass{article}
\usepackage{spconf,amsmath,graphicx}
\usepackage{setspace}
\usepackage{bibspacing}

\usepackage{amsmath}
\usepackage{newtxmath}
\usepackage{multirow}
\usepackage{cite}
\usepackage{hyperref}
\makeatletter
\newcommand{\thickhline}{%
	\noalign {\ifnum 0=`}\fi \hrule height 1.0pt
	\futurelet \reserved@a \@xhline
}

\ninept
\title{Speaker Identity Preservation in Dysarthric Speech Reconstruction by Adversarial Speaker Adaptation}
\name{Disong Wang$^{1}$, Songxiang Liu$^{1}$, Xixin Wu$^{1}$, Hui Lu$^{1}$, Lifa Sun$^{2}$, Xunying Liu$^{1}$, Helen Meng$^{1,3}$}
\address{$^{1}$The Chinese University of Hong Kong, Hong Kong SAR, China\\$^{2}$SpeechX Limited, Shenzhen, China\\$^{3}$Centre for Perceptual and Interactive Intelligence, Hong Kong SAR, China\\\footnotesize{\{dswang, sxliu, wuxx, luhui, xyliu, hmmeng\}@se.cuhk.edu.hk}, lfsun@speechx.cn}

\begin{document}
%\ninept
%
\maketitle
%
% \vspace{-2em}
\begin{abstract}
Dysarthric speech reconstruction (DSR), which aims to improve the quality of dysarthric speech, remains a challenge, not only because we need to restore the speech to be normal, but also must preserve the speaker’s identity. The speaker representation extracted by the speaker encoder (SE) optimized for speaker verification has been explored to control the speaker identity. However, the SE may not be able to fully capture the characteristics of dysarthric speakers that are previously unseen. To address this research problem, we propose a novel multi-task learning strategy, i.e., adversarial speaker adaptation (ASA). The primary task of ASA fine-tunes the SE with the speech of the target dysarthric speaker to effectively capture identity-related information, and the secondary task applies adversarial training to avoid the incorporation of abnormal speaking patterns into the reconstructed speech, by regularizing the distribution of reconstructed speech to be close to that of reference speech with high quality. Experiments show that the proposed approach can achieve enhanced speaker similarity and comparable speech naturalness with a strong baseline approach. Compared with dysarthric speech, the reconstructed speech achieves 22.3\% and 31.5\% absolute word error rate reduction for speakers with moderate and moderate-severe dysarthria respectively. Our demo page is released here\footnote{Audio samples: \url{https://wendison.github.io/ASA-DSR-demo/}}.
\end{abstract}
\begin{keywords}
Dysarthric speech reconstruction, voice conversion, adversarial speaker adaptation, speaker identity
\end{keywords}
\vspace{-0.5em}
\section{Introduction}
\label{sec:intro}
\vspace{-0.5em}

Dysarthria arises from various neurological disorders including Parkinson’s disease or amyotrophic lateral sclerosis, leading to weak regulation of articulators such as jaw, tongue, and lips \cite{yunusova2008articulatory}. Therefore, the resulting dysarthric speech may be perceived as harsh or breathy with abnormal prosody and inaccurate pronunciation, which degrades the efficiency of vocal communication for dysarthric patients. Attempts have been made to improve the quality of dysarthric speech by using various reconstruction approaches, where voice conversion (VC) serves as a promising candidate \cite{yamagishi2012speech}.

The goal of VC is to convert non-linguistic or para-linguistic factors such as speaker identity \cite{mohammadi2017overview}, prosody \cite{rentzos2003transformation}, emotion \cite{aihara2012gmm} and accent \cite{liu2020end}. VC has also been widely applied in reconstructing different kinds of impaired speech including esophageal speech \cite{doi2010esophageal,serrano2019parallel}, electrolaryngeal speech \cite{nakamura2012speaking,kobayashi2018electrolaryngeal}, hearing-impaired speech \cite{biadsy2019parrotron} and dysarthric speech \cite{yamagishi2012speech}, where rule-based and statistical VC approaches have been investigated for dysarthric speech reconstruction (DSR). Rule-based VC tends to apply manually designed, speaker-dependent rules to correct phoneme errors or modify temporal and frequency features to improve intelligibility \cite{rudzicz2011acoustic,kumar2016improving}. Statistical VC automatically maps the features of dysarthric speech to those of normal speech, where typical approaches contain Gaussian mixture model \cite{kain2007improving}, non-negative matrix factorization  \cite{aihara2012consonant,aihara2013individuality}, partial least squares \cite{aihara2017phoneme}, and deep learning methods including sequence-to-sequence (seq2seq) models \cite{wang2020end,doshi2021extending,chen21w_interspeech} and gated convolutional networks \cite{chen2020enhancing}. Though significant progress has been made, previous work generally ignores speaker identity preservation, which loses the ability for patients to demonstrate their personality via acoustic characteristics. Preserving the identities for dysarthric speakers is very challenging since their normal speech utterances are difficult to collect. A few studies \cite{huang2021preliminary,wang2021learning} use a speaker representation to control the speaker identity of reconstructed speech, where the speaker encoder (SE) proposed in our previous work \cite{wang2021learning} is trained on a speaker verification (SV) task by using large-scale normal speech. However, the SE may fail to effectively extract speaker representations from previously unseen dysarthric speech, which lowers the speaker similarity of reconstructed speech.  

% \vspace{-0.5em}
This paper proposes an improved DSR system based on \cite{wang2021learning} by using adversarial speaker adaptation (ASA). The DSR system in \cite{wang2021learning} contains four modules: (1) A \textit{speech encoder} extracting accurate phoneme embeddings from dysarthric speech to restore the linguistic content; (2) A \textit{prosody corrector} inferring normal prosody features that are treated as canonical values for correction; (3) A \textit{speaker encoder} producing a single vector as speaker representation used to preserve the speaker identity; and (4) A \textit{speech generator} mapping phoneme embeddings, prosody features and speaker representation to reconstructed mel-spectrograms. The speaker encoder and speech generator are independently trained by using large-scale normal speech data. We term the resulting integrated DSR system using SV-based speaker encoder as the SV-DSR, which can generate the reconstructed speech with high intelligibility and naturalness. To better preserve the identity of the target dysarthric speaker during speech generation, speaker adaptation can be used to fine-tune the speaker encoder by using the dysarthric speech data. However, this approach inevitably incorporates dysarthric speaking patterns into the reconstructed speech. Hence, we propose to use ASA to alleviate this issue, and the resulting DSR system is termed as the ASA-DSR. For each dysarthric speaker, ASA-DSR is first cloned from SV-DSR and then adapted in a multi-task learning manner: (1) The primary task performs speaker adaptation to fine-tune the speaker encoder by using the dysarthric speech data to enhance the speaker similarity; (2) The secondary task performs adversarial training to alternatively optimize the speaker encoder and a system discriminator, by min-maximizing a discrimination loss to classify whether the mel-spectrograms are reconstructed by ASA-DSR or SV-DSR, which forces the reconstructed speech from ASA-DSR to have a distribution close to that of SV-DSR without dysarthric speaking patterns, rendering the reconstructed speech from ASA-DSR to maintain stable prosody and improved intelligibility.

The main contribution of this paper is the use of proposed ASA approach to effectively preserve speaker identities of dysarthric patients after the reconstruction, without using patients' normal speech that is nearly impossible to collect. It is noted that our work is different from \cite{meng2019adversarial} that aims to achieve robust speech recognition, as the proposed ASA here is used to obtain regularized mel-spectrograms for generating high-quality speech with enhanced speaker similarity.

\vspace{-0.5em}
\section{Baseline approach: SV-DSR}
\vspace{-0.5em}

\begin{figure}[t]
  \centering
  \centerline{\includegraphics[width=\columnwidth]{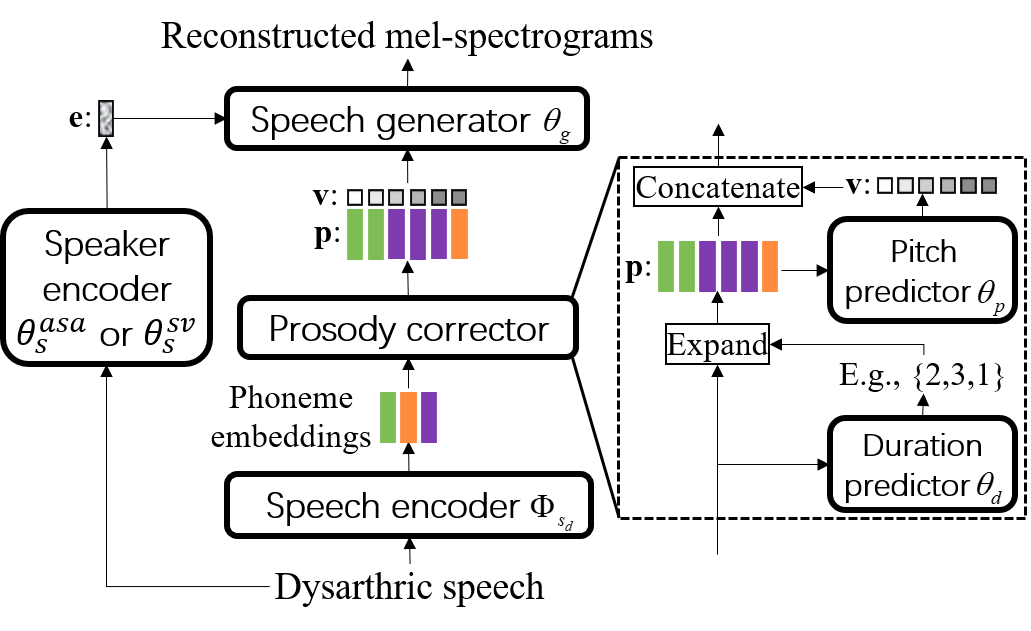}}
  \vspace{-1.0em}
  \caption{The architecture for the SV-DSR system. The ASA-DSR has the same architecture, except that the speaker encoder $\theta_{s}^{sv}$ of SV-DSR is trained for a SV task on the normal speech, while $\theta_{s}^{asa}$ of ASA-DSR is first initialized from $\theta_{s}^{sv}$ and then fine-tuned by the dysarthric speech via proposed ASA.}\label{VC}
  \vspace{-1.5em}
\end{figure}

As shown in Fig. \ref{VC}, our previously proposed SV-DSR system \cite{wang2021learning} contains four modules: speech encoder, prosody corrector, speaker encoder and speech generator. The first three modules respectively produce phoneme embeddings, prosody values and speaker representation; and the fourth module, the speech generator, maps these features to reconstructed mel-spectrograms.

\textbf{Speech encoder:} To recover the content, a seq2seq-based speech encoder is optimized by two-stage training to infer the phoneme sequence: (1) Pre-training on large-scale normal speech data; (2) Fine-tuning on the speech of a certain dysarthric speaker $s_d$ to achieve accurate phoneme prediction. The outputs of pre-trained speech encoder $\Phi _{p}$ or fine-tuned speech encoder $\Phi _{{s_d}}$ are used as phoneme embeddings that denote phoneme probability distributions.

\textbf{Prosody corrector:} As abnormal duration and pitch are two essential prosody factors that contribute to dysarthric speech \cite{kain2007improving}, a prosody corrector is used to amend the abnormal prosody to a normal form, it contains two predictors to respectively infer normal phoneme duration and pitch (i.e., fundamental frequency ($F_0$)). The prosody corrector is trained by a healthy speaker's speech with normal prosodic patterns: (1) Given the phoneme embeddings extracted by the speech encoder $\Phi_p$ as inputs, the phoneme duration predictor $\theta_{d}$ is trained to infer the normal phoneme durations that are obtained from force-alignment via Montreal Forced Aligner toolkit \cite{mcauliffe2017montreal}; (2) The ground-truth phoneme durations are used to align phoneme embeddings and $F_0$ as shown in Fig. \ref{VC}, the expanded phoneme embeddings are denoted as \textbf{p} and fed into the pitch predictor $\theta_{p}$ to infer normal $F_0$ that is denoted by \textbf{v}. The prosody corrector is expected to take in phoneme embeddings extracted from dysarthric speech to infer normal values of phoneme duration and $F_0$, which can be used as canonical values to replace their abnormal counterparts for generating the speech with normal prosodic patterns.

\textbf{Speaker encoder:} The speaker encoder, $\theta _{s}^{sv}$, is trained on a SV task to capture speaker characteristics. $\theta _{s}^{sv}$ takes in mel-spectrograms \textbf{m} of one utterance with arbitrary length to produce a single vector as speaker representation: $\textbf{e}={f_s}(\textbf{m};{{\theta}_{s}^{sv}})$. Following the training scheme in \cite{liu2020end}, $\theta _{s}^{sv}$ is optimized to minimize a generalized end-to-end loss \cite{wan2018generalized} by using normal speech data that is easily acquired from thousands of healthy speakers. 

\textbf{Speech generator:} The speech generator with parameters ${\theta}_g$ predicts mel-spectrograms as: $\textbf{z}={f_g}(\textbf{p},\textbf{v},\textbf{e};{{\theta}_{g}})$. To generate normal speech, the speech generator is trained by using normal speech data from a set of healthy speakers $\mathcal{S}$. Each speaker ${s_i}\sim\mathcal{S}$ has the training data set ${\mathcal{T}_{s_i}}=\{(\textbf{m}_j, \textbf{p}_j, \textbf{v}_j)\}$, where each sample corresponds to one utterance and contains mel-spectrograms $\textbf{m}_j$, expanded phoneme embeddings $\textbf{p}_j$ and pitch features $\textbf{v}_j$. Then speech generator is optimized by minimizing the generation loss $\mathcal{L}_{gen}^{sv}$, i.e., the L2-norm between the predicted mel-spectrograms $ {{\textbf{z}}_{j}^{sv}}$  and $\textbf{m}_j$:
% \vspace{-0.5em}
\begin{equation}
    \mathcal{L}_{gen}^{sv}=\mathbb{E}_{\substack{s_i\sim \mathcal{S}, 
    (\mathbf{m}_{j}, \mathbf{p}_{j}, \mathbf{v}_{j})\sim \mathcal{T}_{s_i}}} \left\| {{\textbf{z}}_{j}^{sv}} - {{\textbf{m}}_{j}} \right\|_{2}
\end{equation}
\begin{equation}
    {{\textbf{z}}_{j}^{sv}} = {{f}_{g}}\left( {{\mathbf{p}}_{j}},{{\mathbf{v}}_{j}}, {{\mathbf{e}}_{j}^{sv}};{{\theta }_{g}} \right), {{\mathbf{e}}_{j}^{sv}}={{f}_{s}}\left( {{\textbf{m}}_{j}};\theta _{s}^{sv} \right)
\end{equation}

During the reconstruction phase, the SV-DSR system takes in the dysarthric speech of speaker $s_d$ to generate reconstructed mel-spectrograms as ${f_g}(\widetilde{\textbf{p}},\widetilde{\textbf{v}},\textbf{e}^{sv};{{\theta}_{g}})$, where $\widetilde{\textbf{p}}$ are phoneme embeddings extracted by fine-tuned speech encoder $\Phi _{{s_d}}$ and expanded with predicted normal duration, $\widetilde{\textbf{v}}$ is predicted normal pitch, and $\textbf{e}^{sv}$ is the speaker representation. Then Parallel WaveGAN (PWG) \cite{yamamoto2020parallel} is adopted as the neural vocoder to transform ${f_g}(\widetilde{\textbf{p}},\widetilde{\textbf{v}},\textbf{e}^{sv};{{\theta}_{g}})$ to speech waveform. SV-DSR is a strong baseline as it can generate the speech with high intelligibility and naturalness. However, the speaker encoder is trained on normal speech, which limits its generalization ability to previously unseen dysarthric speech. Therefore, $\textbf{e}^{sv}$ cannot effectively capture identity-related information of dysarthric speakers. Our experiments found that SV-DSR may even change the gender of speech after the reconstruction.

\vspace{-0.5em}
\section{Proposed approach: ASA-DSR}
\vspace{-0.5em}

\begin{figure}[t]
  \centering
  \centerline{\includegraphics[height=6.5cm]{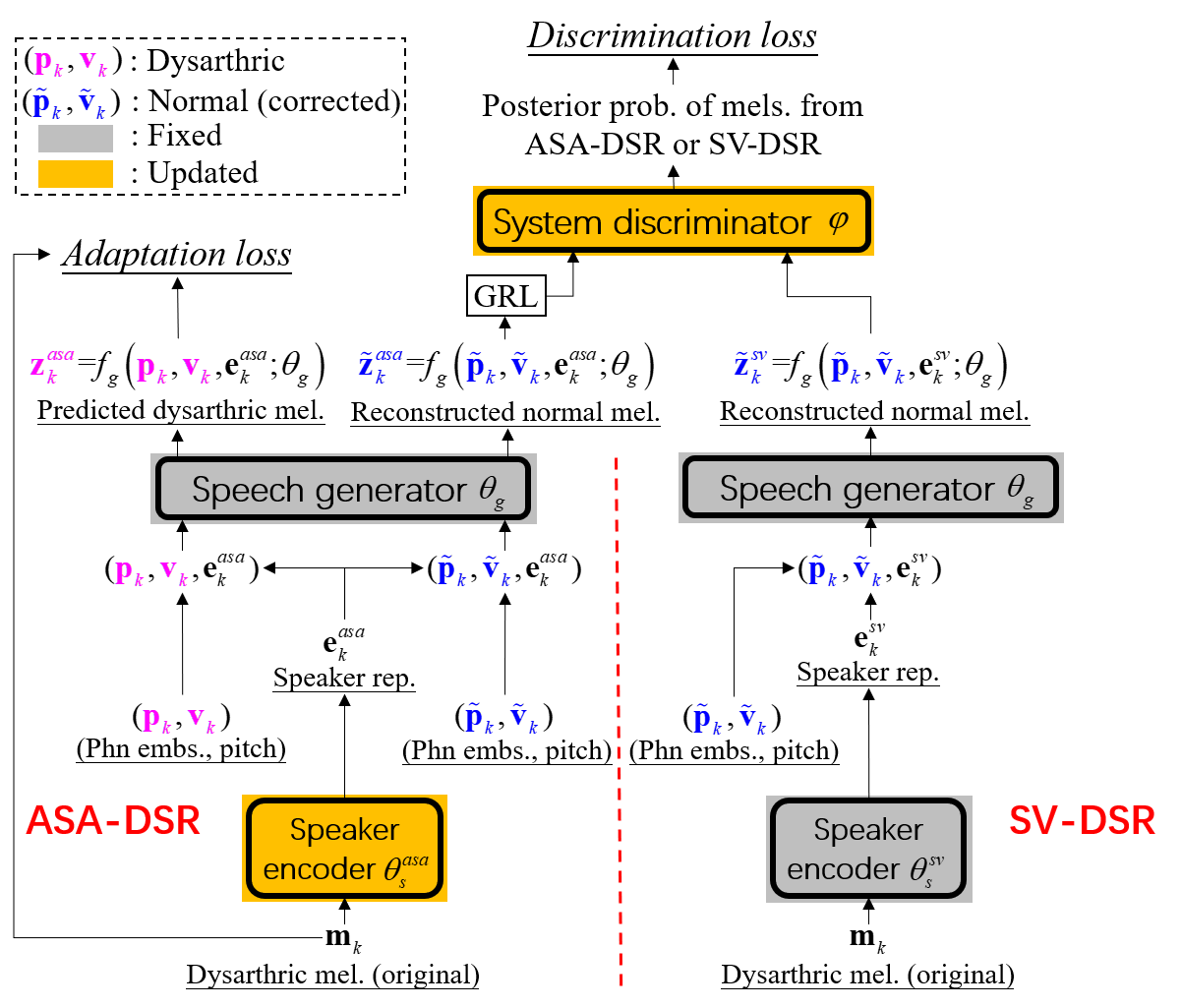}}
  \vspace{-1em}
  \caption{Diagram of ASA. $\textbf{m}_k$ is the mel-spectrogram of dysarthric speech. $\textbf{p}_k$ is phoneme embedding expanded with dysarthric duration, $\textbf{v}_k$ is the pitch of dysathric speech, their normal counterparts are $\tilde{\textbf{p}}_k$ and $\tilde{\textbf{v}}_k$ obtained via prosody corrector. GRL is gradient reversal layer that passes the data during forward propagation and inverts the sign of gradient during backward propagation. Only parameters of ${\theta}_{s}^{asa}$ and $\varphi$ are updated during the ASA process.}\label{asa}
  \vspace{-1.5em} 
\end{figure}

The proposed approach of adversarial speaker adaptation (ASA), as illustrated in Fig. \ref{asa}, aims to enhance speaker similarity, resulting in the proposed ASA-DSR system that shares the same modules as SV-DSR \textit{except for the speaker encoder}. First, ASA-DSR is cloned from SV-DSR, then a system discriminator $\varphi$ is introduced to determine whether its input mel-spectrograms are reconstructed by SV-DSR or ASA-DSR systems. Given a dysarthric speaker $s_d$ with the adaptation data set ${\mathcal{T}_{s_d}}=\{(\textbf{m}_k, \textbf{p}_k, \textbf{v}_k)\}$, where each element corresponds to one dysarthric utterance, $\textbf{p}_k$ are phoneme embeddings extracted by $\Phi _{{s_d}}$ and expanded with dysarthric duration, $\textbf{v}_k$ is dysarthric pitch, their normal counterparts can be obtained via the prosody corrector as $\tilde{\textbf{p}}_k$ and $\tilde{\textbf{v}}_k$, respectively. SV-DSR and ASA-DSR generate reconstructed mel-spectrograms as $\tilde{\textbf{z}}_{k}^{sv}$ and $\tilde{\textbf{z}}_{k}^{asa}$ respectively:
% \vspace{-0.1em}
\begin{equation}
    {\tilde{\textbf{z}}_{k}^{sv}}={{f}_{g}}( {\tilde{\mathbf{p}}_{k}},{\tilde{\mathbf{v}}_{k}}, {{\mathbf{e}}_{k}^{sv}};{{\theta }_{g}}), \;
    {\tilde{\textbf{z}}_{k}^{asa}}={{f}_{g}}( {\tilde{\mathbf{p}}_{k}},{\tilde{\mathbf{v}}_{k}}, {{\mathbf{e}}_{k}^{asa}};{{\theta }_{g}})
\end{equation}
where ${{\mathbf{e}}_{k}^{sv}}$ and ${{\mathbf{e}}_{k}^{asa}}$ are respectively produced from the speaker encoders ${\theta}_s^{sv}$ (from SV-DSR) and  ${\theta}_s^{asa}$ (from ASA-DSR) to control the speaker identity. Besides, ASA-DSR predicts dysarthric mel-spectrograms as ${\textbf{z}}_{k}^{asa}$ used for adaptation:
% \vspace{-0.1em}
\begin{equation}
    {{\textbf{z}}_{k}^{asa}}={{f}_{g}}( {{\mathbf{p}}_{k}},{{\mathbf{v}}_{k}}, {{\mathbf{e}}_{k}^{asa}};{{\theta }_{g}}) 
\end{equation}
Then speaker encoder ${\theta}_{s}^{asa}$ of ASA-DSR and discriminator $\varphi$ are alternatively optimized with remaining networks frozen. On one hand, $\varphi$ is optimized to minimize the discrimination loss $\mathcal{L}_{dis}$:
% \vspace{-0.1em}
\begin{equation}
    {{\mathcal{L}}_{dis}}={{\mathbb{E}}_{({{\textbf{m}}_{k}},{{\textbf{p}}_{k}},{{\textbf{v}}_{k}})\sim {\mathcal{T}_{{{s}_{d}}}}}}\left\{ \mathcal{L}_{dis}^{sv}+\mathcal{L}_{dis}^{asa} \right\}
\end{equation}
\begin{equation}
    {\mathcal{L}}_{dis}^{sv}=\log \left(1-{{f}_{d}}\left(\tilde{\textbf{z}}_{k}^{sv};\varphi \right)\right), {\mathcal{L}}_{dis}^{asa}=\log {{f}_{d}}\left(\tilde{\textbf{z}}_{k}^{asa};\varphi \right)
\end{equation}
where $f_d(*;\varphi)$ is the posterior probability of mel-spectrograms reconstructed by SV-DSR. On the other hand, ${\theta}_{s}^{asa}$ is optimized to minimize the multi-task learning (MTL) loss $\mathcal{L}_{MTL}$:
% \vspace{-0.1em}
\begin{equation}
    {{\mathcal{L}}_{MTL}}={{\mathbb{E}}_{({{\textbf{m}}_{k}},{{\textbf{p}}_{k}},{{\textbf{v}}_{k}})\sim {\mathcal{T}_{{{s}_{d}}}}}}\left\{ \mathcal{L}_{adapt}-\lambda \mathcal{L}_{dis} \right\}
\end{equation}
\begin{equation}
    \mathcal{L}_{adapt}=\left\| {{\textbf{z}}_{k}^{asa}}-{{\textbf{m}}_{k}} \right\|_{2}
\end{equation}
where $\lambda$ is set to 1 empirically. The primary task minimizes the adaptation loss $\mathcal{L}_{adapt}$ to force speaker encoder ${\theta}_{s}^{asa}$ to effectively capture speaker characteristics from the dysarthric speech, so that enhanced speaker similarity can be achieved in reconstructed mel-spectrograms $\tilde{\textbf{z}}_{k}^{asa}$. The secondary task maximizes the discrimination loss $\mathcal{L}_{dis}$ to force $\tilde{\textbf{z}}_{k}^{asa}$ to have a similar distribution to $\tilde{\textbf{z}}_{k}^{sv}$ that has high intelligibility and naturalness, which facilitates  $\tilde{\textbf{z}}_{k}^{asa}$ to maintain normal pronunciation patterns as $\tilde{\textbf{z}}_{k}^{sv}$ . As a result, the proposed ASA-DSR preserves the capacity of SV-DSR to reconstruct high-quality speech, while achieving improved capacity for preserving the speaker identity of the target dysarthric speaker $s_d$.

\vspace{-0.5em}
\section{Experiments}
\vspace{-0.5em}
\subsection{Experimental Settings}
The datasets used in our experiments contain LibriSpeech \cite{panayotov2015librispeech}, VCTK \cite{veaux2016superseded}, VoxCeleb1 \cite{nagrani2017voxceleb}, VoxCeleb2 \cite{chung2018voxceleb2}, LJSpeech \cite{ito2017lj} and UASPEECH \cite{kim2008dysarthric}. Speech encoder $\Phi_p$ is pre-trained by 960h training data of LibriSpeech, prosody corrector is trained by the data of a healthy female speaker from LJSpeech, speaker encoder $\theta_{s}^{sv}$ is trained by Librispeech, VoxCeleb1 and VoxCeleb2 with around 8.5K healthy speakers, speech generator $\theta_{g}$ and PWG vocoder are trained by VCTK. For dysarthric speech, two male speakers (M05, M07) and two female speakers (F04, F02) are selected from UASPEECH, where M05/F04 and M07/F02 have moderate and moderate-severe dysarthria respectively. We use the speech data of blocks 1 and 3 of each dysarthric speaker for fine-tuning speech encoder and ASA, and block 2 for testing. The inputs of speech encoder are 40-dim mel-spectrograms appended with deltas and delta-deltas which results in 120-dim vectors, the targets of speech generator are 80-dim mel-spectrograms, all mel-spectrogtams are computed with 400-point Fourier transform, 25ms Hanning window and 10ms hop length. $F_0$ is extracted by the Pyworld toolkit\footnote{\url{https://github.com/JeremyCCHsu/Python-Wrapper-for-World-Vocoder}} with the 10ms hop length. To stabilize the training and inference of $F_0$ predictor, we adopt the logarithmic scale of $F_0$. All acoustic features are normalized to have zero mean and unit variance.

The speech encoder, prosody corrector, speaker encoder and speech generator adopt the same architectures as in \cite{wang2021learning}, where the speaker encoder contains 3-layer 256-dim LSTM followed by one fully-connected layer to obtain the 256-dim vector that is L2-normalized as the speaker representation \cite{liu2020end}. The pre-training and fine-tuning of speech encoder are performed by Adadelta optimizer \cite{zeiler2012adadelta} with 1M and 2K steps respectively by using learning rate of 1 and batch size of 8. Both duration and $F_0$ predictors are trained by Adam optimizer \cite{kingma2014adam} with 30K steps by using learning rate of 1e-3 and batch size of 16, speech generator is optimized in a similar way except that the training steps are set to 50K. The training of speaker encoder by using normal speech follows the scheme in \cite{liu2020end}. Convolution-based discriminator of StarGAN \cite{choi2018stargan} is used as the system discriminator and alternatively trained with the speaker encoder during ASA for 5K steps. Four DSR systems are compared: (1) SV-DSR; (2) ASA-DSR; (3) SA-DSR, which is an ablation system that performs speaker adaptation similar with ASA-DSR but without adversarial training; and (4) E2E-VC \cite{wang2020end}, which is an end-to-end DSR model via cross-modal knowledge distillation, where the speaker encoder used in SV-DSR is added to control the speaker identity. 

\begin{table}[t]
  \caption{Comparison Results of MOS with 95\% Confidence Intervals for Speaker Similarity.}
  \label{similarity}
%   \vspace{-1em}
  \centering
  \scalebox{0.97}{
  \begin{tabular}{c|c|c|c|c}
    %\thickhline
    \hline\hline
    Approaches & M05 & F04 & M07 & F02 \\
    \hline
    Original & 4.93$\pm$0.01 & 4.89$\pm$0.02 & 4.95$\pm$0.01 & 4.96$\pm$0.01 \\ \hline
    E2E-VC & 2.66$\pm$0.12 & 2.50$\pm$0.13 & 2.47$\pm$0.16 & 2.27$\pm$0.14 \\
    SV-DSR & 2.70$\pm$0.14 & 2.27$\pm$0.10 & 2.55$\pm$0.14 & 1.88$\pm$0.13 \\
    SA-DSR & 3.26$\pm$0.09 & 3.04$\pm$0.12 & \textbf{3.25$\pm$0.15} & \textbf{2.99$\pm$0.15} \\
    ASA-DSR & \textbf{3.27$\pm$0.10} & \textbf{3.16$\pm$0.15} & 3.20$\pm$0.13 & 2.93$\pm$0.15 \\
    %\thickhline
    \hline\hline
  \end{tabular}}
  \vspace{-1.5em}
\end{table}

% \vspace{-0.5em}
\subsection{Experimental Results and Analysis}

\subsubsection{Comparison Based on Speaker Similarity}
Subjective tests are conducted to evaluate the speaker similarity of reconstructed speech, in terms of 5-point mean opinion score (MOS, 1-bad, 2-poor, 3-fair, 4-good, 5-excellent) rated by 20 subjects for 20 utterances randomly selected from each of four dysarthric speakers, and the scores are averaged and shown in Table \ref{similarity}. For E2E-VC and SV-DSR that use the SV-based speaker encoder to control the speaker identity, lower speaker similarity is achieved. Through our listening tests, the gender of reconstructed speech by E2E-VC and SV-DSR may be changed especially for female speakers, this shows the limited generalization ability of the SV-based speaker encoder to extract effective speaker representations from the dysarthric speech. However, with the speaker adaptation to fine-tune the speaker encoder, both SA-DSR and ASA-DSR can accurately preserve the gender with improved speaker similarity, showing the necessity of using dysarthric speech data to fine-tune the speaker encoder to effectively capture identity-related information of dysarthric speech.

\begin{table}[t]
  \caption{Comparison Results of MOS with 95\% Confidence Intervals
for Speech Naturalness.}
  \label{naturalness}
%   \vspace{-1em}
  \centering
  \scalebox{0.97}{
  \begin{tabular}{c|c|c|c|c}
    %\thickhline
    \hline\hline
    Approaches & M05 & F04 & M07 & F02 \\
    \hline
    Original & 2.37$\pm$0.08 & 2.49$\pm$0.09 & 1.95$\pm$0.10 & 1.79$\pm$0.09 \\ \hline
    E2E-VC & 3.64$\pm$0.11 & 3.40$\pm$0.13 & 3.58$\pm$0.12 & 3.35$\pm$0.12 \\
    SV-DSR & \textbf{3.88$\pm$0.11} & \textbf{3.92$\pm$0.10} & \textbf{3.80$\pm$0.10} & \textbf{3.79$\pm$0.09} \\
    SA-DSR & 3.56$\pm$0.09 & 3.22$\pm$0.14 & 3.67$\pm$0.11 & 3.38$\pm$0.12\\
    ASA-DSR & 3.84$\pm$0.09 & 3.86$\pm$0.12 & 3.79$\pm$0.09 & 3.75$\pm$0.11 \\
    %\thickhline
    \hline\hline
  \end{tabular}}
  \vspace{-1.5em}
\end{table}

\vspace{-0.8em}
\subsubsection{Comparison Based on Speech Naturalness}
Table \ref{naturalness} gives the MOS results of naturalness of original or reconstructed speech from different systems. We can see that all DSR systems improve the naturalness of original dysarthric speech, and SV-DSR achieves highest speech naturalness scores for all speakers, which shows the effectiveness of explicit prosody correction to generate
the speech with stable and accurate prosody. By using the speaker adaptation without adversarial training, SA-DSR achieves lower naturalness improvements, due to partial dysarthric pronunciation patterns incorporated into the reconstructed speech. This issue can be effectively alleviated by using the proposed ASA to align the statistical distributions of reconstructed speech from ASA-DSR and SV-DSR, which facilitates ASA-DSR to generate high-quality speech that achieves comparable naturalness with SV-DSR.

\begin{table}[t]
  \caption{WER($\Delta$) (\%) Results Comparison, Where '$\Delta$' Denotes the WER Reduction of Different Approaches Compared with Original Dysarthric Speech.}
  \label{asr}
%   \vspace{-1em}
  \centering
  \scalebox{0.9}{
  \begin{tabular}{c|c|c|c|c}
    %\thickhline
    \hline\hline
    Approaches & M05 & F04 & M07 & F02 \\
    \hline
    Original & 91.0 & 81.7 & 95.6 & 95.9 \\
    % PWG & 96.8(5.8$\uparrow$) & 92.2(10.5$\uparrow$) & 97.9(2.3$\uparrow$) & 98.2(2.3$\uparrow$) \\
    \hline
    E2E-VC & 69.8(21.2) & 69.3(12.4) & 73.1(22.5) & 72.0(23.9) \\
    SV-DSR & \textbf{61.7(29.3)} & \textbf{64.6(17.1)} & \textbf{62.7(32.9)} & \textbf{65.3(30.6)} \\
    SA-DSR & 69.6(21.4) & 70.0(11.7) & 67.8(27.8) & 67.2(28.7) \\
    ASA-DSR & 62.5(28.5) & 65.6(16.1) & \textbf{62.7(32.9)} & 65.8(30.1) \\
    %\thickhline
    \hline\hline
  \end{tabular}}
  \vspace{-1.5em}
\end{table}

\vspace{-0.8em}
\subsubsection{Comparison Based on Speech Intelligibility}
Objective evaluation of speech intelligibility is conducted by using a publicly released speech recognition model, i.e., Jasper \cite{li2019jasper}, to test the word error rate (WER) with greedy decoding, and the results are shown in Table \ref{asr}. Compared with original dysarthric speech, SV-DSR achieves largest WER reduction for all dysarthric speakers, showing the effectiveness of prosody correction to improve the speech intelligibility. Compared with SV-DSR, the adaptation version of SV-DSR without adversarial training, i.e., SA-DSR, has smaller WER reduction, which is caused by the incorporation of dysarthric speaking characteristics into reconstructed speech. However, with the proposed ASA to alleviate this issue, ASA-DSR outperforms E2E-VC and SA-DSR and matches the performance of SV-DSR, leading to 22.3\% and 31.5\% absolute WER reduction on average for speakers M05/F04 and M07/F02 that have moderate and moderate-severe dysarthria respectively.

\begin{figure}[t]
  \centering
  \centerline{\includegraphics[width=\columnwidth]{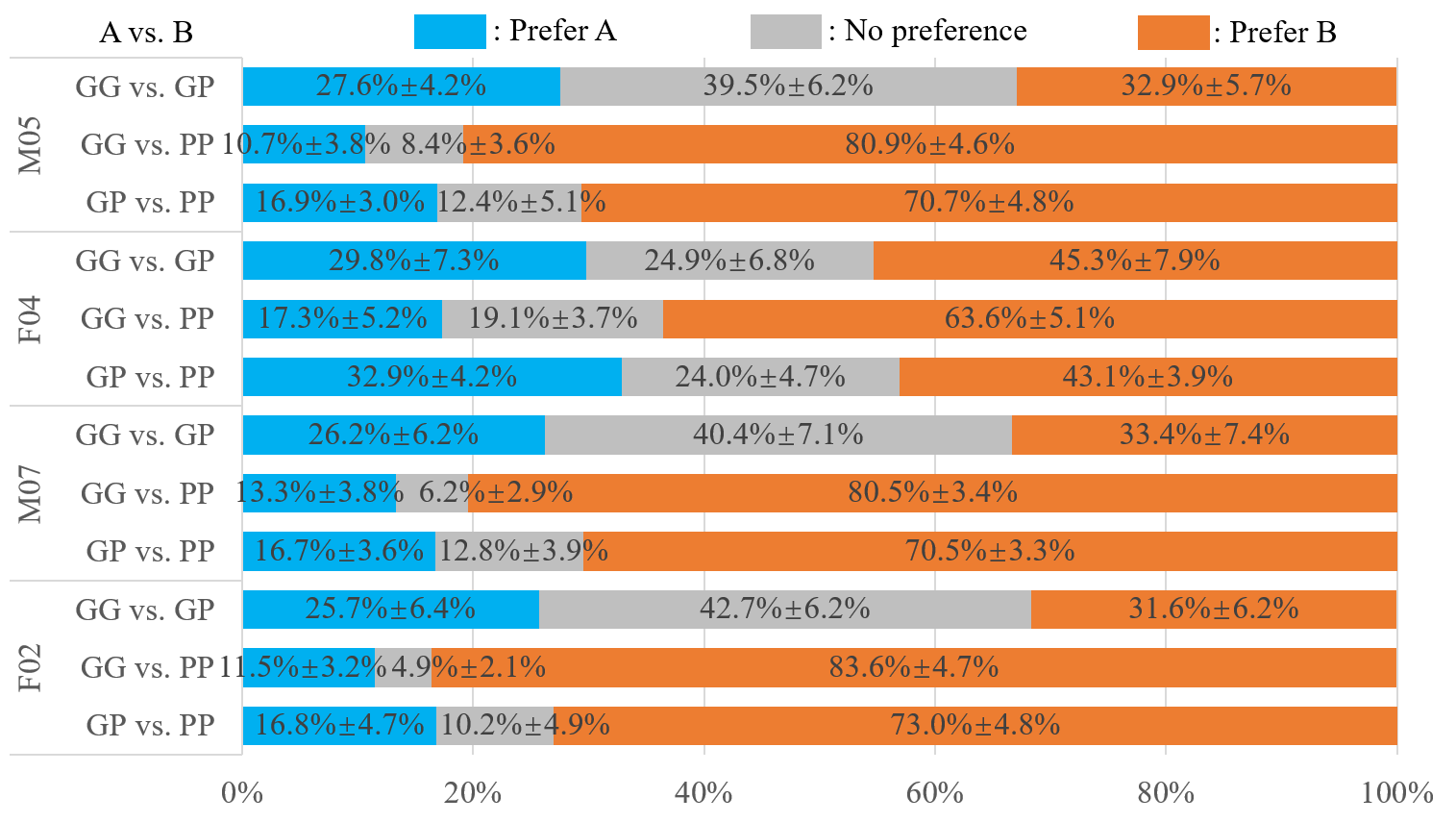}}
  \vspace{-1.0em}
  \caption{AB preference test results with 95\% confidence intervals for different combinations of phoneme duration and $F_0$, where `GG' denotes Ground-truth duration and Ground-truth $\textit{F}_0$, `GP' denotes Ground-truth duration and Predicted $\textit{F}_0$, and `PP' denotes Predicted duration and Predicted $\textit{F}_0$.}\label{AB}
  \vspace{-1.5em}
\end{figure}

\vspace{-0.5em}
\subsubsection{Influence of Phoneme Duration and $F_0$}
We also conduct an ablation study to investigate how the phoneme duration and $\textit{F}_0$ influence the quality of reconstructed speech by the proposed ASA-DSR system. Three combinations of phoneme duration and $\textit{F}_0$ are used to generate the speech. We perform AB preference tests, where listeners are required to select the utterance that sounds more normal, i.e., more stable prosody and precise articulation, from two utterances generated by two different combinations. The results are illustrated in Fig. \ref{AB}. For the comparison ‘GG vs. GP’ (i.e., Ground-truth duration and $\textit{F}_0$ versus Ground-truth duration and Predicted $\textit{F}_0$) of different speakers, more reconstructed speech samples are favored by using predicted normal $\textit{F}_0$ (p-values $\ll$ 0.05). For the comparison `GP vs PP’ (i.e., Ground-truth duration and Predicted $\textit{F}_0$ versus Predicted duration and $\textit{F}_0$), using the predicted normal duration can significantly improve speech quality especially for speakers M05, M07 and F02 who have abnormally slow speaking speed. This shows that both phoneme duration and $\textit{F}_0$ affect speech normality, and the prosody corrector in ASA-DSR derives normal values of phoneme duration and $\textit{F}_0$, which facilitate the reconstruction of speech to have normal prosodic patterns.

\vspace{-0.5em}
\section{Conclusions}
\vspace{-0.5em}

This paper presents a DSR system based on a novel multi-task learning strategy, i.e., ASA, to simultaneously preserve the speaker identity and maintain high speech quality. This is achieved by a primary task (i.e., speaker adaptation) to facilitate the speaker encoder to capture speaker characteristics from the dysarthric speech, and a secondary task  (i.e., adversarial training) to avoid the incorporation of dysarthric speaking patterns into reconstructed speech. Experiments show that the proposed ASA-DSR can effectively achieve dysarthria reductions with improved naturalness and intelligibility, while speaker identity can be effectively maintained with 0.73 and 0.85 absolute MOS improvements of speaker similarity over the strong baseline SV-DSR, for speakers with moderate and  moderate-severe dysarthria respectively. 

\vspace{-0.5em}
\section{ACKNOWLEDGEMENTS}
\vspace{-0.5em}

This research is supported partially by the HKSAR Research Grants Council's General Research Fund  (Ref Number 14208817) and also partially by the Centre for Perceptual and Interactive Intelligence, a CUHK InnoCentre. 

% References should be produced using the bibtex program from suitable
% BiBTeX files (here: strings, refs, manuals). The IEEEbib.bst bibliography
% style file from IEEE produces unsorted bibliography list.
% -------------------------------------------------------------------------
\footnotesize
\bibliographystyle{IEEEbib}
\bibliography{strings,refs}

\end{document}